\documentclass[aps,prl,twocolumn,notitlepage,showpacs,floatfix,superscriptaddress]{revtex4-1}
\usepackage{listings}
\usepackage{bigints}
\usepackage{dcolumn}
\usepackage{tabularx}
\usepackage{bm}
\usepackage{soul}
\usepackage{amsmath,amssymb,graphicx}
\usepackage{lipsum}
\usepackage[colorlinks=true,citecolor=blue,urlcolor=blue,linkcolor=blue]{hyperref}
\usepackage{environ}

\NewEnviron{eqnsplit}{
\begin{equation}
\begin{split}
  \BODY
\end{split}
\end{equation}
}
\DeclareMathOperator{\sinc}{sinc}

\begin{document}
\setlength{\parskip}{0pt}
\title{Electrically Active Domain Wall Magnons in Layered van der Waals Antiferromagnets}
\author{Mohammad Mushfiqur Rahman}
\email{rahman62@purdue.edu}
\affiliation{Elmore Family School of Electrical and Computer Engineering, Purdue University, West Lafayette, IN 47907}
\author{Avinash Rustagi}
\affiliation{Elmore Family School of Electrical and Computer Engineering, Purdue University, West Lafayette, IN 47907}
\author{Yaroslav Tserkovnyak}
\affiliation{Department of Physics and Astronomy, University of California, Los Angeles, CA  90095}
\author{Pramey Upadhyaya}
\email{prameyup@purdue.edu}
\affiliation{Elmore Family School of Electrical and Computer Engineering, Purdue University, West Lafayette, IN 47907}
\affiliation{Purdue Quantum Science and Engineering Institute, Purdue University; West Lafayette, Indiana 47907, USA.}
\affiliation{Quantum Science Center; Oak Ridge, Tennessee 37831 USA.}
\date{\today}

\begin{abstract}
We study theoretically domain wall (DW) magnons|elementary collective excitations of magnetic DWs| in easy-axis layered van der Waals (vdW)  antiferromagnets, where they behave as normal modes of coupled spin superfluids. We uncover that, due to spin-charge coupling in vdW magnets, such DW magnons can be activated by voltage-induced torques, thereby providing a path for their low-dissipation and nanoscale excitation. 
Moreover, the electrical activation and the number of DW magnons at a frequency can be controlled by applying symmetry-breaking static magnetic field, adding tunability of signal transmission by them. Our results highlight that domain walls in vdW magnets provide a promising platform to route coherent spin information for a broad range of explorations in spintronics and magnetism. 
\end{abstract}

\maketitle

\textit{Introduction}|van der Waals (vdW) magnets have recently emerged as an attractive platform to study magnetism in the atomic two-dimensional (2D) limit \cite{gong2017discovery, huang2017layer, sivadas2018stacking, klein2019enhancement, xu2021emergence, jiang2018controlling, jiang2018electric, huang2018electrical, gong2019two}. They offer access to a variety of magnetic phases which can be controlled by tuning: number of layers \cite{huang2017layer}, stacking order \cite{sivadas2018stacking, klein2019enhancement}, Moir\'e potentials \cite{xu2021emergence, hejazi2020noncollinear, wang2020stacking, cheng2022electrically}, charge environment \cite{jiang2018controlling, jiang2018electric, huang2018electrical}, and proximity effects \cite{gong2019two}. This has set the stage to discover magnetic phenomena unique to vdW magnets, along with creating previously unavailable functionalities for spintronics.

A phenomena of central importance in magnetism and spintronics is the coupling of charge degree of freedom with magnons, the (quanta of) elementary collective excitation of magnets \cite{chumak2015magnon}. 
In particular, magnon modes that are attracting significant recent interest are domain wall (DW) magnons| wave-like excitations confined and propagating along topological defects separating differently oriented magnetic regions \cite{winter1961bloch, garcia2015narrow}. Such DW magnons provide model systems to explore spin superfluidity \cite{kim2017magnetic}, enable imaging of non-collinear phases \cite{flebus2018proposal, *flebus2019entangling}, and owing to nanoscale channeling \cite{garcia2015narrow, henry2019unidirectional}, reconfigurability \cite{wagner2016magnetic, albisetti2018nanoscale} and the possibility of transmission along arbitrary-shaped waveguides \cite{garcia2015narrow} offer promising building blocks to wire a broad range of magnon-based classical and quantum circuits \cite{wagner2016magnetic, lan2015spin}. 

So far, coupling charges with the DW magnons has, however, remained challenging. Instead, DW magnons are typically excited by difficult to localize and power-hungry oscillating magnetic fields \cite{matsukura2015control}. This limits the range of experimentally accessible DW magnon wavelengths, as well as, undermines the advantages of low-dissipation propagation and nanoscale confinement offered by them. 
%

In this work, we propose and study theoretically the electrical excitation of DW magnons in layered vdW antiferromagnets (AFM). Leveraging the coupling between spin and charge degree of freedom, a generic feature of the atomic layer thick vdW magnets \cite{jiang2018controlling, huang2018electrical, cenker2020direct}, we show that DW magnons in vdW magnets are \textit{electrically active}; oscillating voltage, by coupling directly with the mode variables, applies torque on DW magnons and triggers their dynamics. Motivated by recent intense interest in layered AFMs \citep{huang2018electrical, jiang2018controlling, cenker2020direct, abdul2020quantum,  sun2019rational, otrokov2019prediction}, we illustrate this in the model system of bilayer Ising vdW AFM. 

Interestingly, we find that the DW magnons in this case behave as the normal modes of coupled DW spin superfluids, which exhibit several unique characteristics when compared to typically studied DW magnons in ferromagnets \cite{garcia2015narrow, henry2019unidirectional, kim2017magnetic}. First, there exists two gapless modes, corresponding to string-like displacement of the DW position and superfluid-like transport of spin current by the DW angle, respectively. Each of these modes disperse linearly as a consequence of the Goldstone theorem. Moreover, exploiting the distinct profile and symmetries of these modes we show that their electrical activity can be selectively turned on and off via static external magnetic fields. This adds the attractive feature of tunability to the signal transmission by them. Finally, taking advantage of the local nature of the electrical activation, we also demonstrate the excitation of DW magnons with nanoscale wavelengths.

Our work extends the scope of unique functionalities offered by vdW magnets to routing of coherent spin signals. The proposed voltage-induced torques could also be used to electrically probe other exotic dynamic magnetic phenomena in insulating vdW magnets, such as the presence of Moir\'e magnons \citep{wang2020stacking}. 

\begin{figure}[hbtp]
\centering
\includegraphics[width=0.49\textwidth]{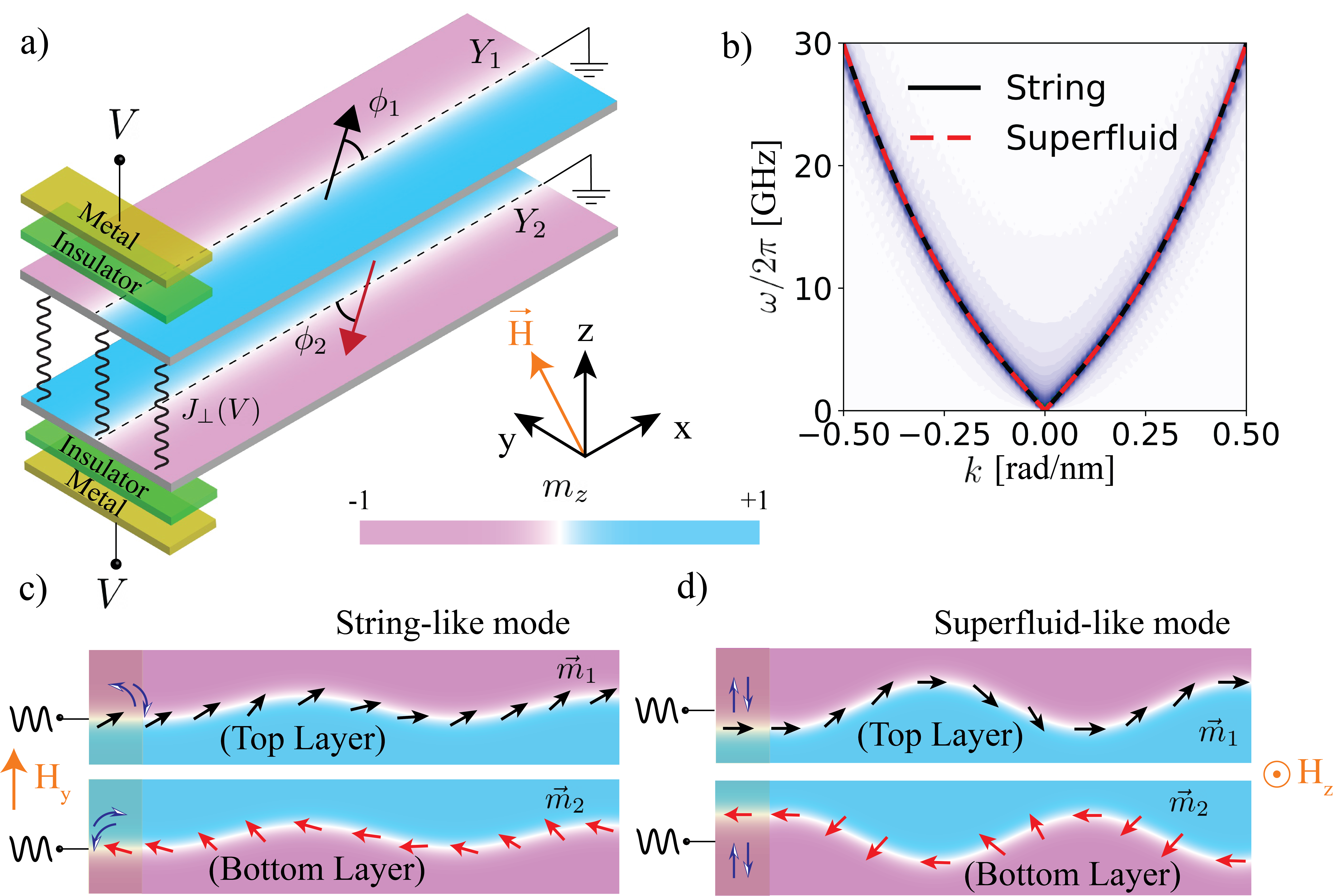}
\caption{\label{Schematic_Structure_Modes} 
(a) Schematic of the proposed geometry for spin wave excitation scheme in the DWs of a bilayer vdW AFM via capacitive coupling to electrical gates. Black and red arrows represent DW spins of the top and bottom layers, parameterized by planar angles $\phi_{1,2}$ and normalized positions $y_{1,2}=Y_{1,2}/\lambda$, where $\lambda = \sqrt{A/K}$ is the DW width. The equilibrium configuration of the DWs are sketched by the dashed straight line. 
(b) Calculated dispersions using both analytics (solid and dashed lines) and micromagnetic  \cite{supp_ref} simulations (background) for the string- and superfluid-like modes (see main text for details) in the absence of bias fields.  (c), (d) An illustration of the propagating DW magnetization profile for the string- and superfluid-like modes.}
\end{figure}

\textit{Central scheme}| As a concrete example, we consider an AFM DW hosted within bilayers of easy-axis layered vdW AFM [Fig.~\ref{Schematic_Structure_Modes}(a), with the easy-axis oriented along the $z$ axis]. This AFM DW can be viewed as two intralayer ferromagnetic (FM) DWs coupled antiferromagnetically via an interlayer exchange. Such AFM DWs can either be written electrically \citep{song2019voltage} or form naturally \citep{zhong2020layer, Sass2020}. 
To describe the DW magnons in this system, we begin by first focusing on the limit of zero interlayer exchange coupling. In this case, the DW magnons become the dynamical modes of two decoupled intralayer FM DWs in an easy-axis magnet. For each DW, the spin within the DW region is oriented along an arbitrary angle $\phi$ in the plane orthogonal to the easy axis, which spontaneously breaks the U(1) symmetry associated with rotations of $\phi$ in easy-axis magnets. Consequently, the dynamical mode of such DWs coincides with the U(1) symmetry-restoring Goldstone mode. 
In close analogy with superfluids, this mode transports spin projected along the easy-axis via gradients of $\phi$, and is thus referred to as the superfluid-like mode \cite{kim2017magnetic}. In FM, the DW position $y$ (here we have defined position normalized to the domain wall width), is canonically coupled with $\phi$ \cite{kim2017magnetic}; the superfluid-like mode thus correspond to harmonic spatio-temporal variations of the canonically conjugate pair ($y, \phi$). 

Turning on a finite interlayer exchange coupling hybridizes the superfluid-like modes living within each layer. The corresponding hybridized normal modes can be described by a new set of canonically conjugate pairs ($y_-, \phi_+$) and ($y_+ ,\phi_- $); the difference in the two DW positions $y_- \equiv y_1-y_2$ (proportional to an out-of-plane spin density) acts as a generator of spin rotations in the xy plane, as described by  $\phi_+ \equiv \phi_1 + \phi_2$. On the other hand, the canting of DW spins $\phi_- \equiv \phi_1 - \phi_2$ (proportional to an in-plane spin density) induces spin rotations in the xz plane (equivalent to the translation of the AFM DW). 

In contrast to the FM case, the position of the AFM DW ($y_+$), which spontaneously breaks the translation symmetry, becomes decoupled with spin rotations in the xy plane ($\phi_+$), which spontaneously breaks the U(1) spin-rotation symmetry. Thus, now there are \textit{two} independent Goldstone modes stemming from \textit{two} continuous symmetries broken by the AFM DW ground state. Consequently, bilayers of layered vdW AFM support two types of DW magnons| describing harmonic spatio-temporal variations of ($y_+ ,\phi_-$) (string-like) and ($y_-, \phi_+$) (superfluid-like)| with each mode dispersing linearly in the long-wavelength limit [see Fig.~\ref{Schematic_Structure_Modes}(c) and (d), with mode dispersions (as calculated below) shown in Fig.~\ref{Schematic_Structure_Modes}(b)].

Moreover, the interlayer hybridization in vdW magnets can be tuned \textit{dynamically} by voltage, for example via a gate coupled capacitively to it \citep{huang2018electrical, jiang2018controlling, cenker2020direct} [Fig.~\ref{Schematic_Structure_Modes}(a)]. This novel functionality can activate the superfluid-like and string-like modes electrically as follows. In the presence of a fixed external magnetic field $H$ applied along the $z$ ($y$) axis, the competition between Zeeman and interlayer exchange energy creates finite $y_-$($\phi_-$), whose value is controlled by the strength of interlayer exchange for a given $H$. Electrical modulation of interlayer exchange thus couples $y_-$($\phi_-$) to voltage for $H$ along $z$ ($y$).  Therefore, an ac voltage with the frequency matched with the DW magnons triggers propagating superfluid-like and/or string-like modes depending on the orientation of the external magnetic field.

Demonstrating these electrically active modes, by combining analytics with micromagnetic simulations, and highlighting their advantages for routing coherent information are the main focus of the remainder of this work.

\textit{Model}|We are particularly interested in the low-energy excitations above the equilibrium configuration of an AFM DW oriented, for simplicity, along a straight line (defined as $y=0$; see Fig.~\ref{Schematic_Structure_Modes}). Within the collective coordinate approach \cite{tretiakov2008dynamics, thiele1973steady}, the Hamiltonian capturing such dynamics is given by (see supplementary \cite{supp_ref} for a detailed derivation):
\begin{equation}
\begin{aligned}
\label{Free energy_after_collective_coordinate}
\mathcal{H}_{\rm dw}= \lambda\bigintssss dx \biggl[\sum_{i=1,2}&2A\bigl[{( \partial_x y_i)^2 + (\partial_x\phi_i)^2}\bigr]\\ +J_\perp (y_i^2+\phi_i^2)
&-2J_\perp (y_1y_2+ \phi_1 \phi_2) \biggr]. 
\end{aligned}
\end{equation}
Eq.~(\ref{Free energy_after_collective_coordinate}) includes contributions from intralayer ferromagnetic exchange ($A$) and interlayer exchange coupling ($J_\perp$). We note that, due to the last term in Eq.~(\ref{Free energy_after_collective_coordinate}), a finite $J_\perp$ hybridizes the uncoupled FM DW modes [described by the harmonic variation of ($y_i,\phi_i$)]. As highlighted in the central scheme section, this will give rise to new normal modes: superfluid-like ($y_-, \phi_+$) and string-like ($y_+ ,\phi_- $), which can be seen explicitly from the equation of motion [see Eq.~(\ref{EOM})] as derived next. 

In particular, we are interested in determining the dynamics of the AFM DW in the presence of capacitively coupled gate-voltage drives, external bias magnetic field (in the yz plane) and dissipation. The latter two are added by including Zeeman interaction ($\mathcal{H}_z$) and Rayleigh function ($R$), which can be written in terms of the field variables as \cite{tatara2008microscopic, supp_ref}: $\mathcal{H}_z= -\lambda M_s \int dx \big[\pi H_y (\sin{\phi_1}-\sin{\phi_2}) + 2 H_z(y_1 - y_2)\big]$ and $R = -(\alpha M_s\lambda/2\gamma) \int dx\, \sum_{i=\{1,2\}}[\dot{y}_i^2 + \dot{\phi}_i^2]$, respectively. Here, $H_i$ is the component of magnetic field along the $i$-th direction, $M_s$ is the saturation magnetization, $\alpha$ is the Gilbert damping parameter and $\gamma>0$ is the gyromagnetic ratio.

For electrical driving, motivated by recent demonstrations \cite{huang2018electrical, jiang2018controlling, cenker2020direct}, we focus on the geometry where the charges on the top and bottom layers are controlled by gates, which for simplicity are connected to the same voltage $V$ [see Fig.~(\ref{Schematic_Structure_Modes})]. In this case, as dictated by the structural and time reversal symmetries, the effect of gate voltage drives is captured by adding the spin-charge coupling term \cite{rustagi2020coupled, huang2018electrical, jiang2018controlling, cenker2020direct}: $\mathcal{H}_{\rm me}=\int dxdy~\xi V \vec{m}_1 \cdot \vec{m}_2$. This term when expressed in terms of DW fields becomes:
$\mathcal{H}_{\rm me}= \lambda\bigintssss dx~\xi V [(y_1-y_2)^2+(\phi_1-\phi_2)^2]$, which can be viewed as voltage tunable interlayer exchange  $J_\perp (V) = J_\perp + \xi V$, where $\xi$ parameterizes the strength of electrical modulation \footnote{We note that symmetry-allowed spin-charge coupling can generate a variety of terms beyond modulation of interlayer exchange, such as the so-called magnetoelectric term which couples the difference in layer magnetizations to the electric field. Here, since we have focused on the geometry where top and bottom gates are connected to the same voltage, i.e. electric field =0, such terms are not included. However, our formalism can be trivially generalized to include electric field coupling for future studies.}.  

We next write the equations of motion by utilizing the Poisson bracket relations between the fields \cite{kim2017magnetic}: $\{y_i, \phi_i\} = (-1)^i\delta(x-x')/(2M_s/\gamma)$, which stem from the fact that the total spin pointing along the $z$ direction acts as the generator of spin rotation about the same axis within each layer. When expressed in terms of the superfluid-like and string-like mode variables identified above, the resultant equations can be written in the linear-response limit as \cite{supp_ref}:
\begin{equation}
\label{EOM}
(1+\alpha^2) \, \delta \dot{q}_{\rm {st/sf}}  = M_{\rm {st/sf}}\delta q_{\rm {st/sf}} + \Delta_{\rm {st/sf}}.
\end{equation}
Here, $\delta q_{\rm {st}} \equiv [\delta\phi_- ,\, \delta y_+]^T$ and $\delta q_{\rm {sf}} \equiv [\delta\phi_+ ,\, \delta y_-]^T$ are the collective coordinates representing small deviations of the string-like and superfluid-like mode variables from the equilibrium, $\Delta_{\rm {st}} \equiv \delta \omega_J \phi_-^{eq} [-\alpha ,\,\, 1]^T$, $\Delta_{\rm {sf}} \equiv \delta \omega_J y_-^{eq} [-1, \, -\alpha]^T$, and
\begin{equation}
\begin{split}
\label{EOM2}
M_{\rm {st}} &= \left[\begin{array}{cc}
-\alpha(\omega_J + \omega_{H_y} -\omega_A \partial_x^2) & \omega_A \partial_x^2 \\ 
\omega_J + \omega_{H_y} -\omega_A \partial_x^2 & \alpha\omega_A \partial_x^2 
\end{array} \right] \\
M_{\rm {sf}} &= \left[\begin{array}{cc}
\alpha (\omega_A \partial_x^2 - \omega_{H_y}) & \omega_A \partial_x^2 -\omega_J \\
\omega_{H_y} - \omega_A \partial_x^2 & \alpha (\omega_A \partial_x^2 - \omega_J)
\end{array} \right],
\end{split}
\end{equation}
with $\omega_J \equiv 2\gamma J_\perp /M_s$, $\delta\omega_J \equiv 2\gamma \xi V /M_s$, $\omega_A \equiv 2\gamma A/M_s$, and $\omega_{H_y} \equiv \gamma \pi H_y \phi_-^{eq}/4$. We highlight that the equations for the string-like and superfluid-like mode variables are not coupled, which confirms that these represent the normal modes in the presence of interlayer exchange.

\begin{figure}[hbtp]
\centering
\includegraphics[width=0.49\textwidth]{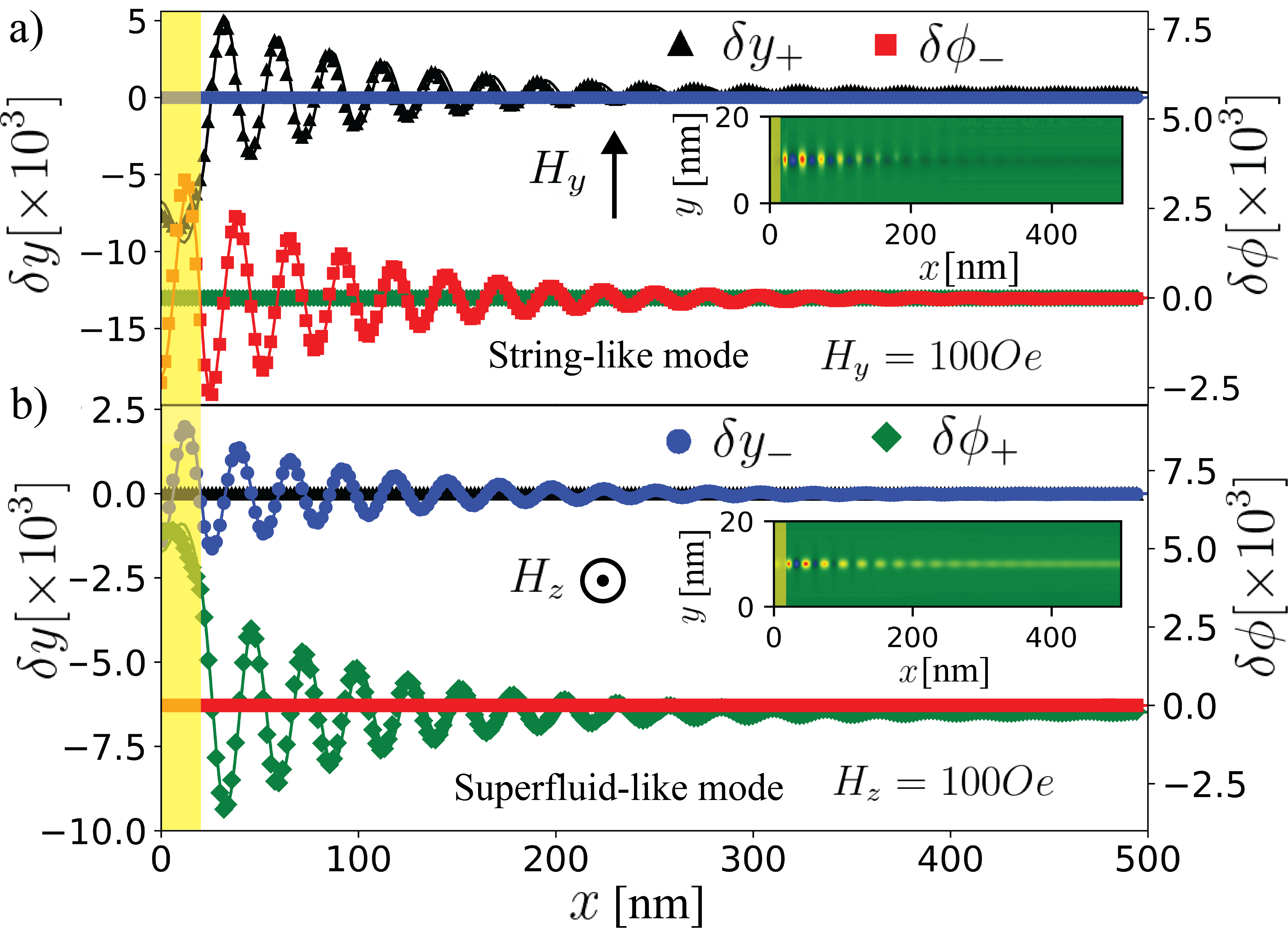}
\caption{\label{FieldDirection_Handle} (a),(b) Spatial profile of DW mode variables  demonstrating their selective excitation via directionally applied fields. \iffalse{}: $H_y$ ($H_z$) excites the string-like (superfluid-like) mode.\fi Theoretical results (solid) are corroborated against micromagnetic simulations (marker). Insets show micromagnetic simulation snapshots of the propagation of dynamic mode variables. \iffalse{} representative dynamic variables, namely $\delta \phi_-(t) \equiv [m_{y1}(t) - m_{y1}(t=0)] + [m_{y2}(t) - m_{y2}(t=0)]$ and $\delta y_-(t) \equiv [m_{z1}(t) - m_{z1}(t=0)] + [m_{z2}(t) - m_{z2}(t=0)]$s. \fi An AC voltage of 4.4 V (equivalent to 10\% change in $J_\perp$) was applied to the gates (marked by yellow bars). The parameters for the layered AFM (CrI$_3$) are: $M_S=$ 1.37$\times10^{-5}$ emu/cm$^2$, $K=$ 0.26 erg/cm$^2$, $A=$ 2.29$\times 10^{-15}$  erg, $\alpha = 0.15$, $J_\perp=$ 0.0354 erg/cm$^2$, and $\xi = -8.06\times10^{-4}$ erg$\mathrm{V^{-1} cm^{-2}}$ \cite{lado2017origin, huang2018electrical, jiang2018controlling}.} 
\end{figure}
\textit{Mode dispersion and Electrical excitation}| The dispersion relations corresponding to the string-like and superfluid-like modes, as obtained by assuming harmonic solutions of Eq.~(\ref{EOM}) in the limit of zero damping and $V=0$, become:
\begin{equation}
\label{dispersions}
\begin{split}
\omega_{st} &=\sqrt{\omega_A k^2(\omega_A k^2 + \omega_J + \omega_{H_y})}\\
\omega_{sf} &=\sqrt{(\omega_A k^2+\omega_J)(\omega_A k^2 + \omega_{H_y})}.
\end{split}
\end{equation}

In particular, we see that in the absence of external magnetic field and in the long-wavelength limit both modes become linear and degenerate with $\omega_{\rm {st/sf}} \approx k\sqrt{\omega_A\omega_J}$, understood as the Goldstone modes corresponding to the spin translational and rotational symmetries. These are plotted in Fig.~\ref{Schematic_Structure_Modes}(b) and corroborated against micromagnetic simulations \cite{supp_ref}. 

%

The terms proportional to $\delta\omega_J(\xi V)$ in the equations of motion act as \textit{voltage-induced torques}, which can be utilized to electrically excite the DW modes. In line with the physical picture presented in the central scheme section, such voltage-induced torques require nonzero $\phi_-^{eq}$ and $y_-^{eq}$ in Eq.~(\ref{EOM}) to activate the dynamics. For a layered antiferromagnet, these can be generated by applying $H_y$ and $H_z$. Therefore, by controlling the direction of external applied static magnetic field a specific mode can be excited electrically. 

To confirm this picture, in Figs.~\ref{FieldDirection_Handle}(a) and (b), we plot \begin{figure}[hbtp]
\centering
\includegraphics[width=0.49\textwidth]{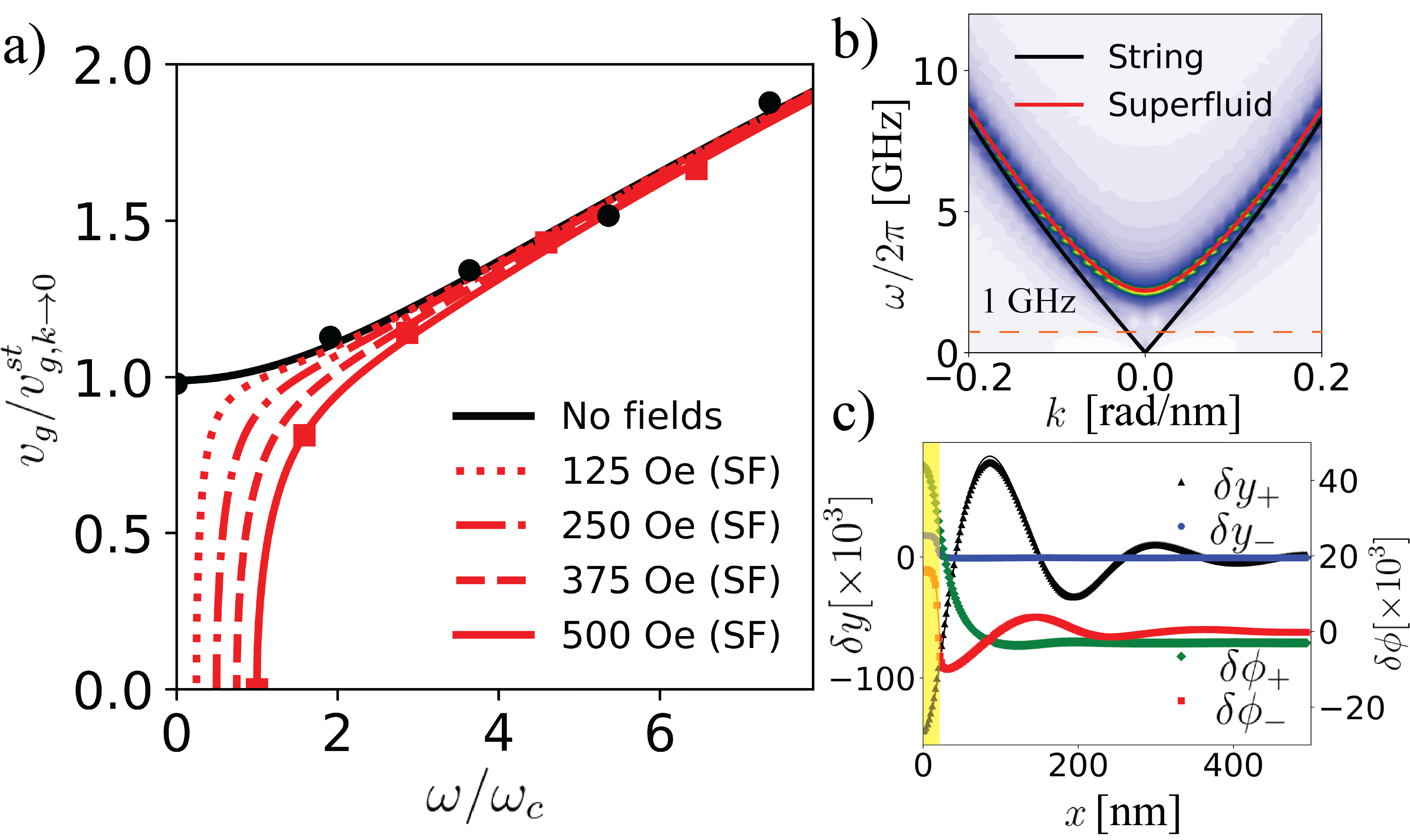}
\caption{\label{speed_and_tunability}  (a) Analytically calculated (solid) and micromagnetic simulations (markers) of group velocity (normalized to $v_{g, k\to 0}^{st/sf} = \sqrt{\omega_A\omega_J}$, the long-wavelength value in the absence of fields) for the string-(blue) and the superfluid-like (red) modes as a function of excitation frequency (normalized to $\omega_c = \sqrt{\omega_J\omega_{H_y}}$, the field-induced gap in superfluid-like mode) for different $H_y$ strengths. (b) Magnon dispersion relations for both the modes at $H_y = 500$ Oe. (c) Dynamic response of the DWs obtained via electrically exciting the system at the gapped frequency (1 GHz) demonstrating the excitation of only the string-like mode variables.}
\end{figure} 
the dynamic response of DWs as obtained by Eq.~(\ref{EOM}), as well as, via micromagnetic simulations \footnote{These are obtained by directly integrating the Landau-Lifshitz-Gilbert (LLG) equations which is equivalent to applying the Hamilton's equations for the Hamiltonian shown in Eq.~(\ref{Free energy_after_collective_coordinate})}. As expected, we note that for a $y$-directed bias field, string-like mode ($\delta y_+, \delta \phi_-$) is excited, while for the field applied along the $z$ axis, superfluid-like mode ($\delta y_-, \delta \phi_+$) is activated. Furthermore, our micromagnetic simulation snapshots (Fig.~\ref{FieldDirection_Handle} insets) confirm that the modes are indeed confined within the DW region. The analytic equations of motion exhibit a good quantitative match with our micromagnetic simulations, demonstrating the proposed scheme of electrical driving captures the underlying physics well. 

\textit{Tunability}| In addition to turning on and off the electrical activity of superfluid-like and string-like modes, external magnetic field can also tune the group velocity and the number of DW magnons available at a frequency by modifying the mode dispersions. For example, an in-plane field can dramatically tune the properties of superfluid-like mode. This is because $H_y$ breaks the spin-rotational symmetry and thereby introduces a gap in the superfluid-like mode dispersion. On the other hand, the string-like mode is left gapless as $H_y$ does not break translation symmetry. 

To study such tunability, we show the group velocity of the modes as a function of drive frequency for various values of $H_y$ [Fig.~\ref{speed_and_tunability}(a)], along with the mode dispersions for an exemplary value of $H_y=500$~Oe [Fig.~\ref{speed_and_tunability}(b)]. We find that the superfluid-like mode acquires a gap, which is given by $\omega_c = \sqrt{\omega_J\omega_{H_y}}$, while the string-like mode remains gapless. Correspondingly, the group velocity of superfluid-like mode shows tunability over a large range, increasing from zero at $\omega=\omega_c$ and approaching the value for string-like mode for large $\omega$. Furthermore, when the applied $H_y$ is such that $\omega_c$ is greater than the frequency of the voltage drive, the transmission of coherent signal is only possible via the string-like channel. In order to demonstrate this, we apply a canted field with $H_y = H_z$ and excite the system with a frequency below $\omega_c$ [Fig.~\ref{speed_and_tunability}(c)]. As both $H_y$ and $H_z$ are non-zero, voltage drives can couple to both string-like and superfluid-like modes. However, our results clearly demonstrate the excitation of only the string-like mode.

\textit{Nanometer-wavelength excitation}|One outstanding challenge in scaling magnon-based circuits to nanoscale dimensions is the ability to efficiently generate short-wavelength magnons \citep{sluka2019emission, van2016tunable, whitehead2017theory, mozooni2015direct, davies2017mapping, whitehead2017theory}. Thanks to the possibility of designing nanoscale gate features, our proposed electrical excitation scheme in vdW magnets provides a route to mitigate this challenge. 

\begin{figure}[hbtp]
\centering
\includegraphics[width=0.49\textwidth]{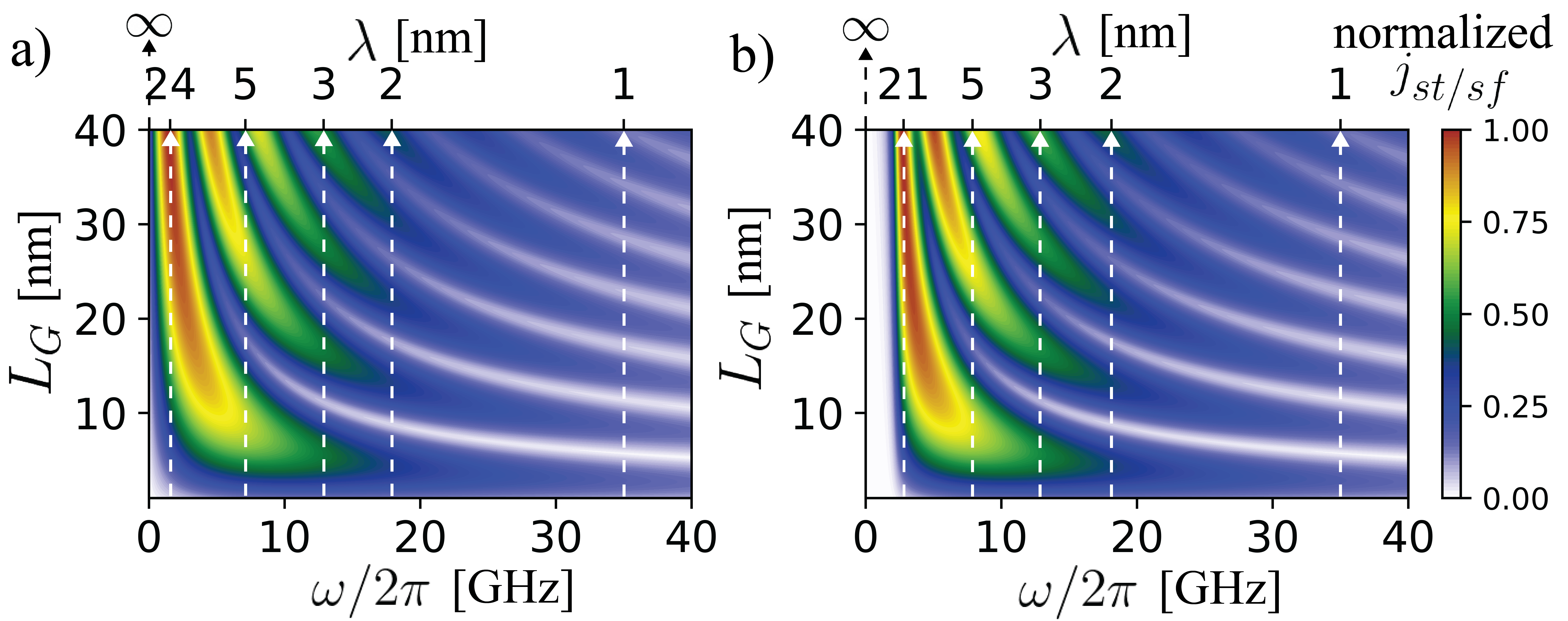}
\caption{\label{spin_current} Analytically calculated spin current amplitudes at a distance $x=100$ nm from the left edge for the (a) string-like ($j_{st}$) and (b) superfluid-like ($j_{sf}$) modes as a function of gate width ($L_g$) and excitation frequency. A canted bias magnetic field with $H_y= H_z = 500$ Oe was applied to excite both the modes.}
\end{figure}

To understand the efficiency of voltage-induced excitation and transport at short wavelengths, we study the effect of gate width ($L_g$) and frequency (equivalently wavelength)  on the ``spin current" at a particular distance away from the gated region (100 nm from the left edge). To this end, from Eq.~(\ref{EOM}), considering $\dot{\delta\phi_-}+\alpha\dot{\delta y_+} -\omega_A \left(\partial_x^2 \delta y_+\right) =0$ and $ \dot{\delta y_-} -\alpha \dot{ \delta \phi_+}+\omega_A \left(\partial_x^2 \delta \phi_+\right) -\omega_{H_y} \delta \phi_+=0$ as the spin continuity relations, the spin currents for the string- and superfluid-like modes are defined as: $j_{st}(x, t)= -2A\partial_x y_+(x, t)$ and $j_{sf}(x, t)= 2A\partial_x \phi_+(x,t)$. We plot in Fig.~\ref{spin_current} the corresponding spin current amplitudes (normalized to the maximum obtained values), which shows an oscillatory dependence on the gate width and frequency before eventually going to zero for high frequency and small gate lengths.

The approach of spin current to zero can be attributed to the reduction of magnon decay length with higher frequency (due to smaller magnon lifetime $\sim 1/\alpha \omega$) and/or eventual decay of the excitation efficiency. The latter is, in turn, proportional to the Fourier component of the source region at the excitation wavelength \citep{mahmoud2020introduction}. For a rectangular gate, the Fourier component is given by a sinc function of $L_g$ and wavelength, which thus results in the observed oscillatory dependence. Importantly, our results suggest that DW magnons across several wavelengths, reaching as low as few nanometers, can be excited with similar efficiencies by choosing the appropriate gate width.

In summary, we propose domain walls in layered vdW 
AFMs as promising candidates for nanoscale routing of coherent information. While we have focused on AFM domain walls in easy-axis vdw magnets, the proposed voltage-induced torques can also provide an efficient means to electrically probe noncollinear phases and the flow of low-energy magnons in stacking DWs \citep{wang2020stacking}, created, for example, via twisting the vdW bilayers to form a Moir\'e magnon crystal. 

\textit{Acknowledgement}|This work was supported by the National Science Foundation through Grant No. ECCS-1810494.\\

\bibliography{final_reference2.bib}
\end{document}


\title{Supplemental Material: Electrically Active Domain Wall Magnons in van der Waals Antiferromagnets}
\author{Mohammad Mushfiqur Rahman}
\email{rahman62@purdue.edu}
\affiliation{Elmore Family School of Electrical and Computer Engineering, Purdue University, West Lafayette, IN 47907}
\author{Avinash Rustagi}
\affiliation{Elmore Family School of Electrical and Computer Engineering, Purdue University, West Lafayette, IN 47907}
\author{Yaroslav Tserkovnyak}
\affiliation{Department of Physics and Astronomy, University of California, Los Angeles, CA  90095}
\author{Pramey Upadhyaya}
\email{prameyup@purdue.edu}
\affiliation{Elmore Family School of Electrical and Computer Engineering, Purdue University, West Lafayette, IN 47907}
\affiliation{Purdue Quantum Science and Engineering Institute, Purdue University; West Lafayette, Indiana 47907, USA.}
\affiliation{Quantum Science Center; Oak Ridge, Tennessee 37831 USA.}
\date{\today}
\maketitle
In this supplemental material, the dynamic equations of motion of the coupled domain walls (DWs) are derived within the collective coordinate approach. We also present the details of micromagnetic simulations for the results presented in the main text. 

\section{Coupled domain wall dynamics}
The Hamiltonian of the layered bilayer antiferromagnet capturing intralayer ferromagnetic exchange ($A$), easy axis anisotropy ($K$, along $z$), Zeeman, and antiferromagnetic interlyer exchange ($J_\perp$) \citep{lado2017origin, rustagi2020coupled} can be written as-
\begin{equation}
\begin{aligned}
\label{Free energy}
\mathcal{H}_m= \bigintssss dxdy \biggl[\sum_{i=1,2}\biggl[{A(\nabla \vec{m}_i)^2} -{K(\vec{m}_i.\hat{z})^2} - {M_s\vec{H}.\vec{m}_i} \biggr]
+J_\perp \vec{m}_1 \cdot \vec{m}_2\biggr],
\end{aligned}
\end{equation}
where $\vec{m_1}$ and $\vec{m_2}$ are the unit vectors oriented along the magnetization fields in layers 1 and 2, respectively and $\vec{H}$ is the externally applied magnetic field. To illustrate the underlying symmetries associated with domain wall position and angles more clearly, it is convenient to parameterize the magnetization vectors with spherical angles. Thus, we write $\vec{m}_i = (\sin\theta_i \cos\psi_i, \sin\theta_i \sin\psi_i, \cos\theta_i)$, where $\theta_i$ and $\psi_i$ are, respectively, the polar and azimuthal angles of the magnetization vector. The dynamics of a system comprising of two antiferromagnetically coupled DWs with ferromagnetic intralayer exchange can be found by solving the Euler-Lagrange equations with the Lagrangian $\mathcal{L} = (M_s/\gamma)\int dxdy \cos{\theta} \dot{\psi} - \mathcal{H}_m$, where the first term originates from the spin Berry phase. The equilibrium configuration associated with this Lagrangian is given by the Walker Ansatz \cite{schryer1974motion}. Thus, for a coupled bilayer system with DWs lying along the $x$ direction, this yields:
\begin{equation}
\label{walker_ansatz}
    \cos\theta_i(y) = (-1)^i \tanh(y/\lambda-y_i),\quad \psi_1(y)=\phi_1,  \quad \textrm{and} \quad \psi_2(y)=\phi_2 + \pi
\end{equation}
where $y_i$ is the position of the DW for the i$^{th}$ layer, $\lambda$ is the DW width, and $\phi_i$ is the deviation of the DW azimuthal angles from its equilibrium (i.e. $\psi_1 = 0$ and $\psi_2 = \pi$) . Thus, upon substituting the Walker Ansatz in Eq.~(\ref{Free energy}) we get:
\begin{equation}
\begin{split}
\label{Free energy_after_collective_coordinate_trigonometric}
\mathcal{H}_{\rm dw} = \lambda \bigintssss dx \bigg[ \sum_{i=1,2}&2A\left[{( \partial_x y_i)^2 +(\partial_x\psi_i)^2}\right] -2M_s H_z (y_1-y_2) - \pi M_s H_y (\sin\phi_1 - \sin\phi_2) \\
&+ 2J_\perp (y_1-y_2) [\coth(y_1-y_2) - \csch(y_1-y_2) \cos(\phi_1-\phi_2) ]\bigg], 
\end{split}
\end{equation}
where we have considered the external field to be oriented in the $yz$ plane. Deviation from such a rigid DW Ansatz is expected to give rise to higher-order bending rigidity effects. However, based on the excellent match between our analytical model and micromagnetic simulations, we conclude that such high-energy effects can be regarded as negligible for the considered parameter range. In addition, throughout this work, we focus on the dynamical regime where the deviations ($y_1 - y_2$) and ($\phi_1 - \phi_2$) are small. Thus, keeping terms up to second order in these deviations, and ignoring the field terms, we get:
\begin{equation}
\begin{aligned}
\label{Free energy_after_collective_coordinate_after_taylor}
\mathcal{H}_{\rm dw}\approx \lambda\bigintssss dx \biggl[\sum_{i=1,2}&2A\left[{( \partial_x y_i)^2 + (\partial_x\phi_i)^2}\right] + J_\perp\left[(y_1-y_2)^2 +(\phi_1-\phi_2)^2\right]],
\end{aligned}
\end{equation} 
which is Eq.~(1) of the main text. In presence of an antiferromagnetic coupling, the Hamiltonian respects two continuous symmetries: the in-phase translation and rotation of DW position and angle, respectively. Thus, we switch to the new coordinates $y_{\pm} = y_1 \pm y_2$ and $\phi_{\pm} = \phi_1 \pm \phi_2$. Moreover, according to the Nambu-Goldstone theorem, spontaneous breaking of these two symmetries give rise to two Goldstone modes each having a linear dispersion relation. Thus, we expect two massless Goldstone modes associated with the dynamic variations of $y_+$ and $\phi_+$. To explain the dynamics associated with these two modes under linear-response, these variables can be broken down into their equilibrium and dynamic components, i.e., $y_{\pm}=y_{\pm}^{eq} + \delta y_{\pm} $ and $\phi_{\pm}=\phi_{\pm}^{eq} + \delta \phi_{\pm}$. Thus, the total free energy up to second order in terms of the collective coordinates can be written as:
\begin{equation}
\begin{split}
\mathcal{H}_{\rm dw} \approx \lambda \int dx \biggl[ A \bigl[ (\partial_x \delta \phi_+)^2 + &(\partial_x \delta y_+)^2 +(\partial_x \delta \phi_-)^2 + (\partial_x \delta y_-)^2 \bigr]+ J_\perp \bigl[ (\delta \phi_-^2 + \delta y_-^2) +2(\phi_-^{eq} \delta \phi_- + y_-^{eq} \delta y_-) \bigr] \\
& -M_s \pi H_Y \bigl[ \delta \phi_- - \frac{\phi_-^{eq}(\delta \phi_+^2 +\delta \phi_-^2 )}{8}\bigr]-2H_Z M_s \delta y_- \biggr],
\end{split},
\end{equation}
Similarly, the spin berry phase term can be expressed as:
\begin{equation}
\begin{split}
\frac{M_s}{\gamma}\int dxdy \cos{\theta} \dot{\psi} = \frac{2 M_s \lambda}{\gamma} \bigintssss dx (\psi_1 \dot{y}_1 - \psi_2 \dot{y}_2)
=\frac{M_s\lambda}{\gamma}\int dx\left[\phi_-\Dot{y_+} + \phi_+\Dot{y_-} -\pi(\Dot{y_+}-\Dot{y_-})\right],
\end{split}
\end{equation}
where $\dot{\mathcal{O}}$ is the notation for time-derivative. We introduce dissipation into our model through the Rayleigh dissipation function as:
\begin{equation}
    R =-\frac{\alpha M_s\lambda}{2\gamma}\int dx\left( 
    \Dot{y_+}^2 + \Dot{y_-}^2 + \Dot{\phi_+}^2 +\Dot{\phi_-}^2\right).\\
\end{equation}
Now, using the Euler-Lagrange equations, the linearized equations of motion for the coupled DWs can be written as: \begin{equation}
\label{string1}
\dot{\delta\phi_-}+\alpha\dot{\delta y_+} -\omega_A \partial_x^2 \delta y_+ =0
\end{equation}
\begin{equation}
\label{string2}
\begin{aligned}
 \dot{\delta y_+}- \alpha \dot{\delta \phi_-} +\omega_A \partial_x^2 \delta \phi_-  -\omega_{\perp}\delta \phi_-& -\frac{\gamma \pi H_Y \phi_-^{eq}}{4}\delta \phi_- = \omega_J \delta \phi_- +\delta \omega_J \phi_-^{eq}
\end{aligned}
\end{equation}
\begin{equation}
\label{superfluid1}
 \dot{\delta\phi_+} +\alpha\dot{\delta y_-} -\omega_A \partial_x^2 \delta y_- = -\omega_J \delta y_- -\delta \omega_J y_-^{eq}
\end{equation}
\begin{equation}
\label{superfluid2}
 \dot{\delta y_-} -\alpha \dot{ \delta \phi_+}+\omega_A \partial_x^2 \delta \phi_+ -\omega_{\perp}\delta \phi_+ -\frac{\gamma \pi H_Y \phi_-^{eq}}{4}\delta \phi_+ =0,
\end{equation}\\
where $\omega_J= 2\gamma J_\perp /M_s$, $\delta\omega_J= 2\gamma \xi V /M_s$, and $\omega_A= 2\gamma A/M_s$. These equations are presented as Eq. (2) of the main text. 
\newpage
\section{Micromagnetic simulations}
The micromagnetic simulations were performed using the open-source GPU-accelerated finite-difference package MuMax3 \cite{vansteenkiste2014design} which numerically integrates the Landau-Lifshitz-Gilbert equation \cite{gurevich2020magnetization}:
\begin{equation}
    \label{LLGeqn}
    \frac{d\vec{m}}{dt} = -\gamma \vec{m}\times\vec{H}_{eff} + \alpha \vec{m}\times \frac{d\vec{m}}{dt},
\end{equation}
where $\vec{H}_{eff}$ is the effective field including contributions from externally applied, intra- and interlayer exchange, and the anisotropy fields. The code uses the Dormand-Prince (RK45) solver with an adaptive time step to integrate the LLG equation. In line with the geometry proposed in the main text, we consider here  the bilayers a layered antiferromagnet whose parameters are summarized in the table below. The cell size has been chosen to be  512 $\times$ 256 per layer with a Neumann boundary condition. The effects of temperature has not been taken into account in the simulations. 
\begin{table}[h]
\caption{\label{tab:table1} Material parameters considered (CrI$_3$).}
\begin{ruledtabular}
\begin{tabular}{lcr}
Parameter & Value\\
\hline
Saturation Magnetization, $M_S$ & 1.37$\times10^{-5}$emu/cm$^2$\\
Easy axis anisotropy, $K$& 0.26 erg/cm$^2$\\
Intralayer Exchange, $A$ & 2.29$\times 10^{-15}$  erg\\
Interlayer Exchange, $J_\perp$ & 0.0354 erg/cm$^2$\\
Damping Constant, $\alpha$ & 0.15\\
Layer thickness, $d$ & 0.4172 nm\\
Length, $L$ & 500 nm\\
Width, $w$ & 20 nm\\
Domain wall width, $\lambda$ & 0.94 nm\\
\end{tabular}
\end{ruledtabular}
\end{table}

To obtain the spin wave spectrum of the magnonic waveguide a procedure similar to Ref \cite{venkat2012proposal} was followed. Throughout the simulations, we considered bilayers of rectangular films with meshes of 512 $\times$ 256 per layer. The equilibrium domain wall configurations were created by the MuMax3 built-in function ``twoDomain()" followed by a relaxation using the built-in ``relex()" and ``minimize()" functions. 
\begin{figure}[hbtp]
\centering
\includegraphics[width=0.999\textwidth]{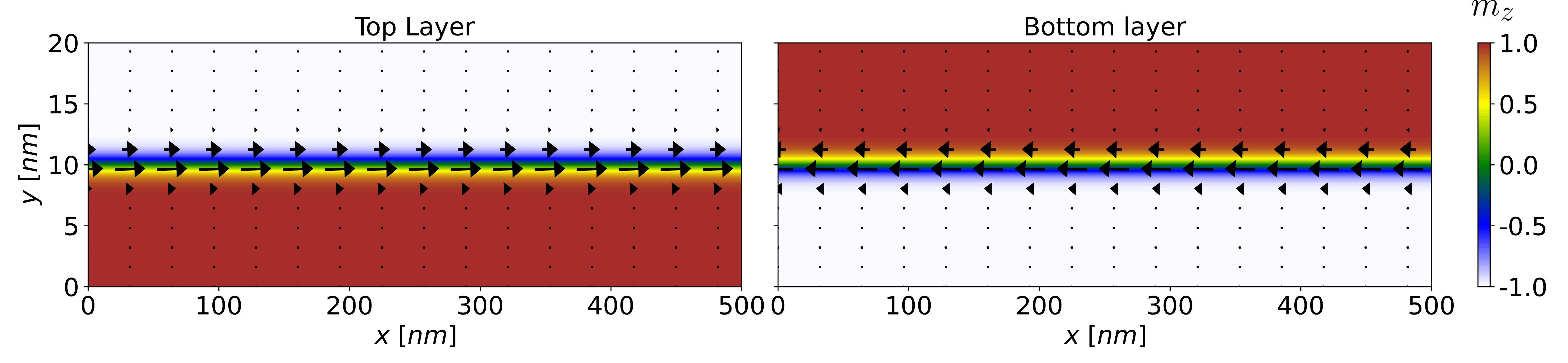}
\caption{\label{DW_static}  Equilibrium configuration of the coupled DWs as obtained from the micromagnetic simulation. The arrows indicate the in-plane magnetization components.}
\end{figure}

 We excite spin waves by adding a small time and space dependent magnetic field of the following form:
\begin{equation}
    \vec{h}(x, y, t) = h_0 \frac{\sin{[k_c(x-L/2)]}}{k_c(x-L/2)} \frac{\sin{[2\pi f_c(t-t_0)]}}{2\pi f_c(t-t_0)} \hat{y}, 
\end{equation}
where $h_0 = $1000 Oe is amplitude of the excitation field, $k_c = $1.25 rad/nm is the cutoff wavevector, $f_c =$ 40 GHz is the cutoff frequency, and a delay time of $t_0 = $1 ns was chosen.  The signal was applied for a total of 10 ns and the dynamic response of the magnetization were recorded. The dispersion relations were calculated by transforming the time-dependent magnetization into the frequency and wavevector space using two-dimensional fast Fourier transformation (FFT). The linewidths of the dispersion relations in the main text is determined by the Gilbert damping parameter as well as the Fourier discretization stepsize.

To excite spin waves electrically, we simulate the effect of voltage by applying a time-varying sinusoidal interlayer exchange $[J_\perp(x, t)]$ between between the layers. The modulation was applied only to the region identified as gates in the main text. 
\begin{equation}
  J_\perp(x, t) =
    \begin{cases}
      J_\perp^0 \sin{(2\pi f t)}, & 0 \leq x \leq L_g\\
      J_\perp^0, & x > L_g
    \end{cases}       
\end{equation}
From the micromagnetic simulations, the domain wall position and angles were extracted by numerically fitting the magnetization profile to Walker Ansatz. 
\bibliography{supp.bib}